\documentclass[pra,amsfonts,showpacs,eqsecnum]{revtex4}              

\usepackage{latexsym}  
\usepackage[]{graphicx}
\usepackage{graphics}
\evensidemargin=\oddsidemargin
\usepackage{times}      
\usepackage{bm}
\def\mcl{\mskip-4mu}
 
\begin{document}

\title{Conditional probabilities and density operators in quantum modeling}
\author{John M. Myers}
\affiliation{Gordon McKay Laboratory, Division of Engineering and Applied
Sciences\\ Harvard University, Cambridge, Massachusetts 02138}

\date{07 June 2005\\[30pt]}

\begin{abstract}   

Based on a recent proof of free choices in linking equations to the
experiments they describe, I clarify relations among some purely
mathematical entities featured in quantum mechanics (probabilities,
density operators, partial traces, and operator-valued
measures), thereby allowing applications of these entities to the
modeling of a wider variety of physical situations.

Conditional probabilities associated with projection-valued measures
are expressed by introducing {\em conditional density operators},
identical in some but not all cases to the usual reduced density
operators. By lifting density operators to the extended Hilbert space
featured in Neumark's theorem, I show an obstacle to extending
conditional density operators to arbitrary positive operator-valued
measures (POVMs); however, tensor products of POVMs are compatible
with conditional density operators.

By way of application, conditional density operators together
with the free choice of probe particles allow the so-called
postulate of state reductions to be replaced by a theorem. A
second application demonstrates an equivalence between one form of
quantum key distribution and another, allowing a formulation of
individual eavesdropping attacks against transmitted-state BB84 to
work also for entangled-state BB84.

\end{abstract}
\pacs{03.65.Ta, 03.67.Dd} 
 
\maketitle

\flushbottom
\section{Introduction}\label{sec:1}

A recent proof confirms what to some will seem a commonplace: the
equations of quantum mechanics are separated by a logical gap from
their application to describing experiments with devices, so that
choosing equations to describe devices involves an irreducible element
of judgment \cite{0404113}.  This finding impacts a core question of
interpretation: what does quantum mechanics describe
\cite{leggett}? Although a wide variety of interpretations
accept the assumptions on which the proof depends, the proof tells us that
quantum mechanics by itself describes nothing, but instead offers a
mathematically articulated language that people speak to describe what
they see or expect or think possible in experiments \cite{ams,JOptB}.

With the recognition of a logical gap between the equations and
experiments comes an opportunity to examine relations among some
purely mathematical entities---probabilities, density operators,
partial traces---separated out from the choices and judgments necessary
to apply them to describing experiments.  Based on this examination, 
I will show uses of {\em conditional density operators} defined in
relation to the trace rule by which quantum mechanics generates
probabilities from density operators and positive operator-valued
measures.

The next section reviews the expression of joint and conditional
probabilities by tensor products of projection-valued measures.
Section \ref{sec:3} generalizes from projection-valued measures to
positive operator-valued measures (POVMs) with a proposition that
lifts density operators to an extended Hilbert space associated with
Neumark's theorem.  Diagrams show how tensor products of POVMs (but
not generic operator products) express joint probabilities, leading to
conditional density operators useful in modeling measurements of
composite systems.

Section~\ref{sec:4} shows how choices in the application of quantum
mathematics to the description of devices allows a single, discrete
measure space to be applied in describing diverse situations.
Section~\ref{sec:5} offers two applications of the conditional
probabilities and partial traces: (1) a demonstration that the use of
reduced states in quantum mechanics requires no postulate about
``effects of a measurement on a state''; and (2) a demonstration of the
equivalence of two forms of quantum key distribution (QKD) \cite{qkd},
which I term ``transmitted-state BB84'' and ``entangled-state BB84''
\cite{BB84} with the result that a known formulation
\cite{slutsky} for studying individual eavesdropping attacks
against transmitted-state BB84 applies also to entangled-state
BB84.

\newpage
\section{Probabilities, Operator-valued measures, and quantum
modeling}\label{sec:2}

By a probability measure I mean a completely additive set function
\cite{kol}, or, in more modern words, a positive measure
\cite{mackey,rudin} of total measure 1.  Let $\Omega$ be a topological
space; let $\mathcal{M}(\Omega)$ be the $\sigma$-algebra of measurable
subsets in $\Omega$, making $\Omega$ into a measurable space. Let
$\mu$ denote any probability measure on a fixed $\mathcal{M}(\Omega)$.
For any measurable sets $X$ and $Y$, with $\mu(Y) > 0$, the
conditional probability of $X$ given $Y$ is defined \cite{kol} as:
\begin{equation}
\mu(X|Y)\stackrel{\rm def}{=}\mu(X\cap Y)/\mu(Y).
\label{eq:conDef}
\end{equation}

Let $\mathcal{B}(\mathcal{H})$ denote the set of bounded, linear
operators on a Hilbert space $\mathcal{H}$.  With
$\mathcal{M}(\Omega)$ as above, a positive operator-valued measure
(POVM) is any function
\begin{equation}M:\mathcal{M}(\Omega) \rightarrow
  \mathcal{B}(\mathcal{H})
\label{eq:MDef}
\end{equation} 
satisfying: (1) $M(\emptyset) = 0$, $M(\Omega) = \bm{1}_{\mathcal{H}}$;
(2) each $M(X)$ is self-adjoint and non-negative; (3) if $X\cap Y = 0$
then $M(X\cup Y) = M(X)+M(Y)$; and (4) for every $|u\rangle,|v\rangle \in
\mathcal{H}$, the set function $M_{u,v}$ defined by $M_{u,v}(X)=
\langle v|M(X)|u\rangle$ is a complex measure on
$\mathcal{M}(\Omega)$.

A projection-valued measure, sometimes called a projective
resolution of the identity \cite{rudin}, is a special case of a
POVM, denoted here by $E$ in place of $M$, that satisfies the
above requirements, and, in addition:
\begin{eqnarray} & &
\mbox{Each $E(X)$ is a self-adjoint projection $($so $E^2(X) =
E(X)${}$)$;}
\\
&& E(X\cap Y) = E(X)E(Y). \label{eq:E2}
\end{eqnarray}
Note that this last property implies
\begin{equation}
[E(X),E(Y] \stackrel{\rm def}{=}E(X)E(Y)-E(Y)E(X) = 0.
\label{eq:commute}
\end{equation}

I take a density operator on $\mathcal{H}$ to be any positive,
self-adjoint trace-class operator $\rho \in \mathcal{B}(\mathcal{H})$
such that $\mbox{Tr}(\rho) = 1$, where $\mbox{Tr}$ denotes the trace
\cite{vN}.  If $\rho$ is a density operator on $\mathcal{H}$, so is
$U\rho U^\dag$ where $U$ is any unitary operator on $\mathcal{H}$.
For any POVM defined for $\mathcal{M}(\Omega)$ and $\mathcal{H}$,
along with any density operator $\rho$ on $\mathcal{H}$ and any
unitary operator $U$ on $\mathcal{H}$, a probability measure on
$\mathcal{M}(\Omega)$ is defined by
\begin{equation}
\mu(\rho,U,M;X) = \mbox{Tr}[U\rho U^\dag M(X)].
\label{eq:form0}
\end{equation}
In the modeling of experiments one is interested in a family of such
measures, corresponding to various choices of $\rho$, $U$, and $M$.
Often one skips the listing of these ``parameters'' and writes
$\Pr(X)$ in place of $\mu(\rho,U,M;X)$.  Feller speaks of
$\Omega$ as a ``sample space'' and of $\mathcal{M}(\Omega)$ as ``the
set of events'' \cite{feller}, inviting us to think of a ``probability
that a sample point falls in an event set X''; and, correspondingly,
to think of $\mu(X|Y)$ as a conditional probability that a sample
point falls in $X$ given that it falls in~$Y$. (Without
drawing a distinction between a sample point and an event, Dirac
\cite{dirac} speaks of a ``result'' while Peres \cite{peres} speaks of
an ``outcome.'') 

In quantum physics we speak of a preparation of a state $\rho$, a time
evolution $U$, and a meas\-urement expressed by~$M$.  Quantum decision
theory \cite{helstrom} adjoins to this story of ``quantum prob\-ability''
a ``classical probability'' in the choice of the state $\rho$ prepared,
making two kinds of events.  To distinguish these kinds, I rename the
event above a {\em measurement event} and join it to a {\em state event}
expressing preparation of the state.  States are usually thought of as
selected from a discrete space of possible states, in which case the
state event can be denoted simply by the state chosen, resulting in a
compound event consisting of a measurement event $X$ (as before)
together with a state event $\rho$.  The marginal probability for
choosing $\rho$ can be denoted $\Pr(\rho)$.  With this
convention, decision problems are formulated in terms of a (joint)
measure
\begin{equation}
\mu(U,M;\rho,X) = \Pr(\rho)\mbox{Tr}[U\rho U^\dag M(X)],
\label{eq:formDec}
\end{equation}
where on the left-hand side of the equation $\rho$ has moved from the
`parameter side' of the semicolon to the `variable side.'  If $U$ and
$M$ are understood, one can write $\Pr(\rho,X)$ in place of
$\mu(U,M;\rho,X)$.  (Models that go further by randomizing the choices
of $M$ and/or $U$ can be found, but not in this report.)

By {\em quantum modeling} I mean stating probabilities in the form of
Eqs.\ (\ref{eq:form0}) or (\ref{eq:formDec}) (together with
probabilities derived from these by Bayes' rule) as mathematical
language by which to ask questions and make statements pertaining to a
set of trials of devices in a laboratory experiment
\cite{0404113,ams}.  Examples of devices are lasers, lenses, and
detectors. In this language a trial is necessarily described as
consisting of ``preparing a state'' and ``measuring a state,'' in some
cases interspersing these with a temporal evolution $U$.

In quantum modeling, one speaks of various conditional probabilities,
conditioned on whatever types of events are expressed in a model;
hence there can be conditioning not only on {\em measurement
events} but also on {\em state events}:
\begin{enumerate}
\item Understanding some $\rho$, $M$, and $U$, one can speak of the
conditional probability $\Pr(X|Y)$ of a sample point being in
measurement event set $X$, given it is in set $Y$.
\item Understanding some $M$ and $U$, one can speak of the conditional
probability of a meas\-urement event $X$, given the state is $\rho$.
For example, from Eq.\ (\ref{eq:formDec}) and the definition of a
conditional probability, we see what in Eq.\ (\ref{eq:form0}) appeared
as a probability becomes in the context of decision theory a species
of conditional probability:
\begin{equation}
\Pr(X|\rho) = \frac{\Pr(\rho,X)}{\Pr(\rho)} = \mbox{Tr}[U\rho U^\dag
M(X)].
\end{equation}
\end{enumerate}
Variations on both of these appear below.

\subsection{Conditional density operators for a single 
projection-valued measure} 

For the rest of this section and all of
Sec.~\ref{sec:3}, I subsume $U\rho U^\dag$ into $\rho$.  By virtue of
Eqs.\ (\ref{eq:conDef}), (\ref{eq:E2}) and (\ref{eq:commute}),
conditional probabilities of a sample point being in one measurement
event given that the point is in another measurement event fit in neatly
with any single projection-valued measure $E$:
\begin{eqnarray}
(\forall\ X, Y \in \mathcal{M}(\Omega)\mbox{ with } \Pr(Y) >0
)\quad \Pr(X|Y) &=& \frac{\mbox{Tr}[\rho E(X)E(Y)]}{\mbox{Tr}[\rho
E(Y)]}=\frac{\mbox{Tr}[E(Y)\rho E(Y)E(X)]}{\mbox{Tr}[\rho
E(Y)]},\qquad\qquad\quad
\label{eq:joint}
\end{eqnarray}
which allows us to define a ``conditional density operator''
\begin{equation}
\rho|_Y \stackrel{\rm def}{=} \frac{E(Y)\rho
E(Y)}{\mbox{Tr}[\rho E(Y)]},
\label{eq:rhoYDef}
\end{equation}
with the property that
\begin{equation}
(\forall\ X)\quad \Pr(X|Y) =
  \mbox{Tr}[\rho|_Y E(X)].
\end{equation}
Similarly one constructs $\rho|_X$ such that
\begin{equation}
(\forall\ X)\quad \Pr(Y|X) =
  \mbox{Tr}[\rho|_X E(Y)].
\label{eq:rhoX}
\end{equation}
It is easy to check that the conditional density operators $\rho|_Y$
and $\rho|_X$ are non-negative, self-adjoint, and have unit trace.
The expression of a conditional probability by a conditional density
operator depends on the expression of a joint probability by a
product of projections, as illustrated by the following commutative
diagram (where commutative means that alternative ways to compose
mappings arrive at the same thing):
\begin{equation}
{\def\arraystretch{1.6}
\begin{tabular}{ccccc} 
$X$ &\ $\longmapsto$\ & $X\cap Y$ &\
\raise5.5pt\hbox{\rotatebox{180}{$\longmapsto$}}\ &
$Y$ \\[-12pt]
\rotatebox{270}{$\longmapsto$}\ &&\rotatebox{270}{$\longmapsto$}\ 
&&\rotatebox{270}{$\longmapsto$}\ \\ 
$E(X)$ &\ $\longmapsto$\ &\ $E(X)E(Y)$\ &\
\raise5.5pt\hbox{\rotatebox{180}{$\longmapsto$}}\ &
$E(Y)$ 
\end{tabular}.}
\label{eq:diag1}
\end{equation}
\begin{center} 
\vskip4pt
Diagram 1:\quad $E(X\cap Y) = E(X)E(Y)$.\end{center}
\vspace{10pt} Although most of the conditional density operators to be
introduced below are constructed using partial traces, $\rho|_Y$ as
defined by Eq.\ (\ref{eq:rhoYDef}) involves no partial trace and so is
not a reduced density operator, which shows that the concept of a
conditional density operator is distinct from the concept of a reduced
density operator.

\textbf{Note}: The conditional probabilities above are just
probabilities that a measurement sample point that falls in a set $Y$
also falls in a set $X$; this has nothing to do with ``consecutive
measurements,'' nor with so-called ``state reductions'' or ``collapse
of a wave function'' postulated by Dirac \cite{dirac} and von Neumann
\cite{vN}.

\subsection{Tensor products of projection-valued measures}

Consider a special case that arises in the modeling of systems
viewed as composites of subsystems $A$ and $B$, so that
\begin{eqnarray}
\Omega & = & \Omega_A \times \Omega_B \quad\mbox{(cartesian product)},
\nonumber \\ \mathcal{H} &\equiv & \mathcal{H}_{AB} = \mathcal{H}_A
\otimes \mathcal{H}_B \quad\mbox{(tensor product)}, \nonumber \\ E&  =& 
E_A
\otimes E_B\quad \mbox{ with } E_A \mbox{ on } \mathcal{H}_A \mbox{ and }
E_B \mbox{ on }\mathcal{H}_B,\nonumber \\ \rho & = & \rho_{AB} \in
\mathcal{B}(\mathcal{H}_{AB}),\nonumber \\ U & = &\bm{1}_{AB}.
\end{eqnarray}
The tensor product of projection-valued measures is defined by:
\begin{equation}
(\forall\ X_A \in \mathcal{M}(\Omega_A),Y_B \in
\mathcal{M}(\Omega_B))\quad (E_A \otimes E_B)(X_A \times X_B) =
E_A(X_A)\otimes E_B(Y_B).\ 
\end{equation}
Then we have
\begin{eqnarray}
(X_A \times \Omega_B) \cap (\Omega_A \times Y_B)&  = &(X_A \times Y_B),
\\[4pt]
(E_A\otimes E_B)[(X_A \times \Omega_B) \cap (\Omega_A \times Y_B)] & =
& (E_A\otimes E_B)(X_A \times Y_B) =
E_A(X_A)\otimes E_B(Y_B),
\label{eq:special}
\end{eqnarray}
and the following diagram commutes:
\begin{equation}
{\def\arraystretch{1.6}
\begin{tabular}{ccccccccc}
$X_A \mcl$&\ $\longmapsto$\ &$\mcl X_A \times\Omega_B
\mcl$&\ $\longmapsto$\ &$(X_A
\times \Omega_B) \cap (\Omega_A \times Y_B) \mcl$& \raise5.5pt\hbox{\rotatebox{180}{$
\longmapsto$}}\ &$\mcl
\Omega_A\times Y_B \mcl$& \raise5.5pt\hbox{\rotatebox{180}{$\longmapsto$}}\ 
&$\mcl Y_B$ \\[-12pt]  
\rotatebox{270}{$\longmapsto$}\ &&\rotatebox{270}{$\longmapsto$}\ 
&&\rotatebox{270}{$\longmapsto$}\ &&\rotatebox{270}{$\longmapsto$}\ 
&&\rotatebox{270}{$\longmapsto$}\ \\
$E_A(X_A) \mcl$&\ $\longmapsto$\ &$E_A(X_A)\otimes \bm{1}_B
\mcl$&\ $\longmapsto$\ &$\mcl E_A(X_A)\otimes E_B(Y_B)
\mcl$& \raise5.5pt\hbox{\rotatebox{180}{$\longmapsto$}}\ &$
\bm{1}_A \otimes E_B(Y_B) \mcl$& \raise5.5pt\hbox{\rotatebox{180}{$\longmapsto$}}\ &$
E_B(Y_B)$ 
\end{tabular}.}
\label{eq:diag2}
\end{equation}
\begin{center}
\vskip6pt
Diagram 2: Tensor product of projection-valued measures.
\end{center}

\vspace{5pt}

\noindent For this specialization of Diagram 1, it follows from Eqs.\
(\ref{eq:joint}) and (\ref{eq:special}) that
\begin{eqnarray}
\Pr(X_A|Y_B) & \equiv & \Pr(X_A \times \Omega_B|\Omega_A \times Y_B)
\nonumber \\[8pt] & = & \frac{\mbox{Tr}_{AB}\{[\bm{1}_A\otimes
E_B(Y_B)]\rho_{AB}[\bm{1}_A\otimes E_B(Y_B)](E_A(X_A)\otimes
  \bm{1}_B)\}}{\mbox{Tr}_{AB}\{\rho_{AB}[\bm{1}_A\otimes E_B(Y_B)]\}},
\label{eq:cond0}
\end{eqnarray}
where the subscript $AB$ on the trace indicates the trace over
$\mathcal{H}_{AB}$.  From this follow two distinct ways to
define a conditional density operator:
\begin{enumerate}
\item Eq.\
(\ref{eq:rhoX}) implies:
\begin{eqnarray} 
\Pr(X_A|Y_B) & = &
\mbox{Tr}_{AB}[\rho^{(1)}_{AB}|_{Y_B}(E_A(X_A)\otimes
\bm{1}_B)],\label{eq:cond1}
\end{eqnarray}
with
\begin{eqnarray}
\rho^{(1)}_{AB}|_{Y_B} & \stackrel{\rm def}{=}&
\frac{[\bm{1}_A\otimes E_B(Y_B)]\rho_{AB}[\bm{1}_A\otimes
E_B(Y_B)]}{\mbox{Tr}_{AB}\{\rho_{AB}[\bm{1}_A\otimes E_B(Y_B)]\}}.
\end{eqnarray}

\item The alternative definition takes
advantage of partial traces to obtain a conditional density operator
as a reduced density operator:
\begin{equation}
\Pr(X_A|Y_B) =  \frac{\mbox{Tr}_A\mbox{Tr}_{B}\{[\bm{1}_A\otimes
E_B(Y_B)]\rho_{AB}[\bm{1}_A\otimes E_B(Y_B)](E_A(X_A)\otimes
  \bm{1}_B)\}}{\mbox{Tr}_{AB}\{\rho_{AB}[\bm{1}_A\otimes E_B(Y_B)]\}},
\end{equation}
which, with $X \equiv X_A$ and $Y \equiv Y_B$ understood, can
be written with less clutter as
\begin{eqnarray}
\Pr(X|Y) &= & \frac{\mbox{Tr}_A\mbox{Tr}_{B}\{[\bm{1}_A\otimes
E_B(Y)]\rho_{AB}[\bm{1}_A\otimes E_B(Y)](E_A(X)\otimes
  \bm{1}_B)\}}{\mbox{Tr}_{AB}\{\rho_{AB}[\bm{1}_A\otimes E_B(Y)]\}}
\nonumber \\ 
 &= &
\mbox{Tr}_A[\rho^{(2)}_A|_Y(E_A(X)\otimes
  \bm{1}_B)],\label{eq:cond2}
\end{eqnarray}
with\vadjust{\eject}
\begin{eqnarray}
\rho^{(2)}_A|_Y &\stackrel{\rm def}{=} & 
\frac{\mbox{Tr}_B\{[\bm{1}_A\otimes
E_B(Y)]\rho_{AB}[\bm{1}_A\otimes
E_B(Y)]\}}{\mbox{Tr}_{AB}\{\rho_{AB}[\bm{1}_A\otimes E_B(Y)]\}}
\nonumber \\ & = & \frac{\mbox{Tr}_B\{\rho_{AB}[\bm{1}_A\otimes
E_B(Y)]\}}{\mbox{Tr}_{AB}\{\rho_{AB}[\bm{1}_A\otimes
E_B(Y)]\}}.\label{eq:cond2end}
\end{eqnarray}
\end{enumerate}
I stretch notation to allow $\mbox{Tr}_A$ to indicate either the
partial trace over the factor $\mathcal{H}_A$ of the tensor-product
Hilbert space $\mathcal{H}_A \otimes \mathcal{H}_B$ or the trace over
just the Hilbert space $\mathcal{H}_A$.  Properties
of the partial trace used in these equations are  reviewed in Appendix A;
for example, Eq.\ (\ref{eq:a10}) assures us that the last expression
in Eq.\ (\ref{eq:cond2end}) is a self-adjoint operator on
$\mathcal{H}_A$.

When this mathematics of tensor products is applied to model a
measurement event viewed as a rectangle $X_A \times Y_B$, it is
convenient to speak of $X_A$ and $Y_B$ separately as pertaining to
{\em components} of the measurement event.
  
\section{Positive operator-valued measures and conditional
probabilities}\label{sec:3}

The diagram (\ref{eq:diag1}) fails for generic POVMs, but the
specialization to tensor-products holds for POVMs.  How this works can
be seen from Neumark's theorem, along with a corollary developed
below.

\subsection{Neumark's theorem}
 
Suppose a Hilbert space $\mathcal{H}$ is a subspace of a Hilbert space
$\mathcal{H}^+$.  Let $E^+:\mathcal{M}(\Omega) \rightarrow
\mathcal{B}(\mathcal{H}^+)$ be any projection-valued measure on
$\mathcal{H}^+$, and let $Q \in \mathcal{B}(\mathcal{H}^+)$ be the orthogonal
(hence self-adjoint) projection on $\mathcal{H}$.  (Although consistency
calls for writing this as $Q^+$, to avoid clutter I write just $Q$.) 
Define a POVM
$(QE^+Q)_\mathcal{H}:\mathcal{M}(\Omega) \rightarrow
\mathcal{B}(\mathcal{H})$ by
\begin{equation}(\forall\ X \in \mathcal{M}(\Omega))\quad
(QE^+Q)_\mathcal{H}(X) = (QE^+(X)Q)_\mathcal{H},
\label{eq:QEQ}
\end{equation}
where $\mathcal{H}$ as a subscript denotes the restriction
of an operator to $\mathcal{H}$.

Neumark \cite{neumark,Ak} proved that all POVMs can be expressed this
way: for any POVM $M$ on any Hilbert space $\mathcal{H}$, there exists
a Hilbert space $\mathcal{H}^+$ containing $\mathcal{H}$ as a subspace,
and there exists a projective resolution of the identity $E^+:\mathcal{M}(\Omega)
\rightarrow \mathcal{B}(\mathcal{H}^+)$ such that $M =
(QE^+Q)_{\mathcal{H}}$.

\subsection{Lifting the trace rule to $\mathcal{H}^+$}
It is instructive to lift the trace rule to the extended Hilbert
space $\mathcal{H}^+$.
When we view $\mathcal{H}$ as a subspace of $\mathcal{H}^+$, we see
any vector $|\psi\rangle \in \mathcal{H}$ as a vector $|\psi\rangle
\oplus |0\rangle^\perp$ in $\mathcal{H}^+$, where
$|0\rangle^\perp$ denotes the 0-vector in
$\mathcal{H}^\perp$. Correspondingly, we view any operator $A \in
\mathcal{B}(\mathcal{H})$ as an operator $A \oplus \bm{0}^\perp \in
\mathcal{B}(\mathcal{H}^+)$, where $\bm{0}^\perp$ is the zero operator
on $\mathcal{H}^\perp$.  With $Q \in \mathcal{B}(\mathcal{H}^+)$ the
orthogonal projection onto $\mathcal{H}$ as above, useful elementary
facts are
\begin{eqnarray}
(\forall\ A,B \in \mathcal{B}(\mathcal{H}))(\forall\ C^+, D^+ \in
\mathcal{B}(\mathcal{H}^+))\qquad&& \nonumber \\
 Q(A\oplus\bm{0}^\perp)Q & = &
A\oplus\bm{0}^\perp, \label{eq:X1} \\ QC^+Q & = &
(QC^+Q)_\mathcal{H}\oplus \bm{0}^\perp,\qquad\qquad\qquad
\label{eq:lemX2}
\\ (A\oplus
\bm{0}^\perp)(B\oplus \bm{0}^\perp)& =& (AB) \oplus \bm{0}^\perp,
\label{eq:X3} \\
\mbox{Tr}^+(A
\oplus \bm{0}^\perp) & = &\mbox{Tr}(A),
\label{eq:lemX4} 
\end{eqnarray}
where $\mbox{Tr}$ denotes the trace on $\mathcal{H}$ and
$\mbox{Tr}^+$ denotes the trace on $\mathcal{H}^+$. From these
equations follows

\noindent\textbf{Lemma}: 
\begin{eqnarray}
\mbox{Tr}^+[(A\oplus \bm{0}^\perp)C^+] & = &\mbox{Tr}^+[Q(A\oplus
\bm{0}^\perp)QC^+] \nonumber \\ & = & 
\mbox{Tr}^+[(A\oplus \bm{0}^\perp)QC^+Q] \nonumber \\ & = &
\mbox{Tr}^+\{(A\oplus
\bm{0}^\perp)[(QC^+Q)_\mathcal{H}\oplus\bm{0}^\perp]\} \nonumber \\& = &
\mbox{Tr}^+[A(QC^+Q)_\mathcal{H}\oplus\bm{0}^\perp] \nonumber \\& = &
\mbox{Tr}[A(QC^+Q)_\mathcal{H}].
\label{eq:lemProp}
\end{eqnarray}
Letting $A$ be $\rho$ and $C^+$ be $E^+(X)$ yields

\vspace{5pt}

\noindent\textbf{Proposition}:
\begin{equation}
(\forall\ \rho)\,(\forall\ X \in \mathcal{M}(\Omega))\quad \Pr(X|\rho)
\equiv\mbox{Tr}[\rho M(X)] =\mbox{Tr}^+[(\rho \oplus
\bm{0}^\perp)E^+(X)].
\label{eq:prop}
\end{equation}
Eq.\ (\ref{eq:prop}) allows the trace rule for any given single POVM
to be lifted to $\mathcal{H}^+$, as illustrated in the commutative
diagram
\begin{equation}
{\def\arraystretch{1.9}
\begin{tabular}{ccccc}
 $E^+(X)$&\hskip-14pt\lower.5em\hbox{\rotatebox{315}{$\longmapsto$}}&&
   \lower.2em\hbox{\rotatebox{225}{$\longmapsto$}} & $(\rho\oplus \bm{0}^\perp)\in
\mathcal{B}(\mathcal{H}^+)$\\[-5pt]
 \tiny$\mskip.3mu-$ &&&&\hskip-36pt\tiny$\mskip.3mu-$ \\[-15.5pt]
$|$&&$\mbox{Tr}^+[(\rho\oplus
\bm{0}^\perp)E^+(X)]$&&\hskip-36pt$|$\\[-16pt] 
$|$&&&&\hskip-36pt$|$\\[-16pt]
$Q,|_\mathcal{H}\
|\hphantom{Q,|_\mathcal{H}\
}$&&$=$&&\hskip-36pt$\hphantom{Q,|_\mathcal{H}\ }|\
Q,|_\mathcal{H}$\\[-16pt]
$|$&&&&\hskip-36pt$|$\\[-13pt]
$\downarrow$&&$\mbox{Tr}[\rho
M(X)]$&&\hskip-36pt$\downarrow$\\[0pt]
$M(X)$&\hskip-14pt\raise.6em\hbox{\rotatebox{45}{$\longmapsto
$}}&&\raise.95em\hbox{\rotatebox{135}{$\longmapsto$}}&\hskip-34pt$
\rho\in \mathcal{B}(\mathcal{H})$\end{tabular}.}\label{eq:diag3}
\end{equation}
\begin{center} Diagram 3: Lifting of the trace rule to $\mathcal{H}^+$
 via Neumark's theorem.
\end{center}

\subsection{Obstacle to expressing conditional probabilities with POVMs}
The desire to extend the diagram engendered by Proposition
(\ref{eq:prop}) to $\Pr(X\cap Y|\rho)$ for the set intersection $X\cap
Y$ encounters the following obstacle:
\begin{equation}
{\def\arraystretch{1.9}
\begin{tabular}{ccccc}
 $E^+(X)$&\hskip-6pt$\longmapsto$\hskip-6pt&$E^+(X\!\cap\!Y)\!=\!
E^+(X)E^+(Y)$&\hskip-6pt
\raise5.5pt\hbox{\rotatebox{180}{$\longmapsto$}}\hskip-6pt& $E^+(Y)$\\[-7pt]
 \tiny$\mskip.3mu-$ &&&&\tiny$\mskip.3mu-$ \\[-15.5pt]
$|$&&\hskip-16pt\raise5.5pt\hbox{\rotatebox{270}{$\longmapsto$}}&&$|$\\[-16pt] 
$|$&&&&$|$\\[-16pt]
$|$&&&&$|$\\[-13pt]
$Q,|_\mathcal{H}\ |\hphantom{Q,|_\mathcal{H}\
}$&&$M(X\cap Y)
\!=\![QE^+(X)E^+(Y)Q]_\mathcal{H}$&&$|$\\[-16pt]
$|$&&&&$|$\\[-16pt]
$|$&&&&$|$\\[-13pt]
$\downarrow$&&$\hskip-16pt\ne$&&$\downarrow$\\[0pt]
$M(X)\!=\![QE^+(X)Q]_\mathcal{H}$&$\!\longmapsto\!$&$M(X)M(Y)\!=\!
[QE^+(X)Q]_\mathcal{H}[QE^+(Y)Q]_\mathcal{H}$&$\!\raise5.5pt\hbox{\rotatebox{180}{$\longmapsto$}}\!
$&$M(Y)\!=\![QE^+(Y)Q]_\mathcal{H}$\end{tabular}.}
\label{eq:diag4}
\end{equation}
\begin{center}
\vskip4pt
 Diagram 4: Obstacle to lifting of product of POVMs.
\end{center}
\noindent Because
\begin{equation}
[QE^+(X)E^+(Y)Q]_\mathcal{H} \bm{\ne}
  [QE^+(X)Q]_\mathcal{H}[QE^+(Y)Q]_\mathcal{H},
\end{equation}
we find that except in uninteresting special cases
\begin{equation}
M(X\cap Y) \ne M(X)M(Y).
\end{equation}
Correspondingly, efforts to define reduced states corresponding
to a measurement modeled by $M(X)$ followed by a measurement modeled
by $M(Y)$ encounter conceptual difficulties, touched on by Braunstein
and Caves \cite{BC} and discussed below in connection with probes.
\vfil

\subsection{Tensor products of POVMs}
The obstacle to operator products of POVMs is no impediment to tensor
products. 
Consider a POVM $M_A:\mathcal{M}(\Omega_A)\rightarrow
\mathcal{B}(\mathcal{H}_A)$ and another POVM
$M_B:\mathcal{M}(\Omega_B)\rightarrow \mathcal{B}(\mathcal{H}_B)$.  By
Neumark's theorem, both of these POVMs can be expressed as
restrictions of projections of projection-valued measures on the
respective extended Hilbert spaces:
$E^+_A:\mathcal{M}(\Omega_A)\rightarrow
\mathcal{B}(\mathcal{H}_A^+)$ and
$E^+_B:\mathcal{M}(\Omega_B)\rightarrow \mathcal{B}(\mathcal{H}_B^+)$,
respectively. Let $Q_A \in \mathcal{B}(\mathcal{H}^+_A)$ be the orthogonal
(hence self-adjoint) projection on $\mathcal{H}_A$, and similarly
$Q_B \in \mathcal{B}(\mathcal{H}^+_B)$. 
Because a tensor product of projections is a projection, Diagram 2 for
the extended Hilbert spaces extends downward, via $Q_A$ and $Q_B$
followed by reductions, to $\mathcal{H}$.  
 The neat thing here is that
\begin{equation}
(Q_A\otimes Q_B)(E^+_A\otimes E^+_B)(Q_A\otimes Q_B)
= (Q_AE^+_AQ_A)\otimes(Q_BE^+_BQ_B),
\end{equation}
the restriction of which to the subspace $\mathcal{H}_A\otimes
\mathcal{H}_B$ is just $M_A\otimes M_B$; i.e. we have
\begin{eqnarray}
M_A\otimes M_B & = & [Q_AE_AQ_A\otimes Q_BE_BQ_B]|_{\mathcal{H}_{BA}}
\nonumber \\ & = & [(Q_A \otimes Q_B)(E_A\otimes E_B)(Q_A\otimes
Q_B)]|_{\mathcal{H}_{BA}}.
\end{eqnarray}
Thus we arrive at the diagram:
\begin{equation}
{\def\arraystretch{1.6}
\begin{tabular}{ccccccccccc}
&$E^+_A$ & $\rightarrow$ & $E^+_A\otimes
\bm{1}_B$ & $\rightarrow$ & $E^+_A\otimes E^+_B$ &$\leftarrow$ & $\bm{1}_A
\otimes E_B^+$ & $\leftarrow$ & $E^+_B$ &\\
$Q_A,|_{\mathcal{H}_A}$ & $\downarrow$ && $\downarrow$ && $\downarrow$ &&
$\downarrow$ && $\downarrow$ & $Q_B,|_{\mathcal{H}_B}$\\ 
& $M_A$ & $\rightarrow$ & $M_A\otimes \bm{1}_B$ & $\rightarrow$ &
$M_A\otimes M_B$ & $\leftarrow$ & $\bm{1}_A \otimes M_B$ & $\leftarrow$ &
$M_B$\vadjust{\kern8pt} &\end{tabular}\ .}
\label{eq:diag5}
\end{equation}
\begin{center}
Diagram 5: Tensor product of POVMs. 
\end{center}

Because this shows
\begin{equation}
(M_A\otimes M_B)[(X_A \times
\Omega_B) \cap (\Omega_A \times Y_B)] = M_A(X_A)\otimes M_B(Y_B),
\end{equation}
Eqs.\ (\ref{eq:cond0})--(\ref{eq:cond2}) hold also for non-projective
POVMs; for instance, Eq.\ (\ref{eq:cond2}) becomes (with the
understanding $X \equiv X_A$ and $Y \equiv Y_B$):
\begin{eqnarray}
\Pr(X|Y) &= & \frac{\mbox{Tr}_{AB}[M_A(X)\otimes
M_B(Y)]\rho_{AB}}{\mbox{Tr}_{AB}\{\rho_{AB}[\bm{1}_A\otimes M_B(Y)]\}}
\nonumber \\[6pt] 
& = &\frac{\mbox{Tr}_A\mbox{Tr}_{B}\{[\bm{1}_A\otimes
M_B(Y)]\rho_{AB}[\bm{1}_A\otimes M_B(Y)](M_A(X)\otimes
  \bm{1}_B)\}}{\mbox{Tr}_{AB}\{\rho_{AB}[\bm{1}_A\otimes M_B(Y)]\}}
\nonumber \\[1pt] 
 &= &
\mbox{Tr}_A[\rho_A|_YM_A(X)],\end{eqnarray}
with\vadjust{\kern-10pt}
\begin{eqnarray}
\rho_A|_Y &\stackrel{\rm def}{=} & 
\frac{\mbox{Tr}_B\{[\bm{1}_A\otimes
M_B(Y)]\rho_{AB}[\bm{1}_A\otimes
M_B(Y)]\}}{\mbox{Tr}_{AB}\{\rho_{AB}[\bm{1}_A\otimes M_B(Y)]\}}
=\frac{\mbox{Tr}_B\{\rho_{AB}[\bm{1}_A\otimes
M_B(Y)]\}}{\mbox{Tr}_{AB}\{\rho_{AB}[\bm{1}_A\otimes M_B(Y)]\}}.
\label{eq:cond3}
\end{eqnarray}

\subsection{Tensor products of families of POVMs}
Instead of considering a single POVM on $\mathcal{H}_A$ and a single
POVM on $\mathcal{H}_B$, we consider now two indexed families of
POVMs, $M_{A,\alpha}$ and $M_{B,\beta}$.  Then Diagram 5 threatens to
become decorated with these indices in a curiously asymmetric way,
because the proof of Neumark's theorem presents $\mathcal{H}_B$ not as
a $\beta$-independent subspace of $\mathcal{H}_B^+$, but as one that
varies with $\beta$.

The following corollary restores symmetry to the distribution of
indices and justifies lifting not just one POVM but any family
of POVMs to a single extended Hilbert space related to the base
space by a single (index-independent) orthogonal projection.

\vspace{8pt}

\noindent\textbf{Corollary to Neumark's theorem}: Given a fixed
Hilbert space $\mathcal{H}$ and a fixed $\sigma$-algebra
$\mathcal{M}(\Omega)$ of measurable sets, together with any family of
POVMs $M_\beta:\mathcal{M}(\Omega)\rightarrow
\mathcal{B}(\mathcal{H})$ (indexed by $\beta$); then there is a single
extended Hilbert space $\mathcal{H}^+$ containing $\mathcal{H}$ as a
subspace and a single $Q \in \mathcal{B}(\mathcal{H}^+$), the orthogonal
projection on $\mathcal{H}$, along with a family of projection-valued
measures $E^+_\beta$ on $\mathcal{H}^+$, such that
\begin{equation}
(\forall\ \beta)(\exists E^+_\beta)\quad M_\beta =
  [QE^+_\beta Q]_{\mathcal{H}}.
\end{equation}
\noindent\textit{Proof sketch}: The proof of Neumark's theorem in
\cite{Ak} (and also in \cite{neumark}) deals with each POVM of a
family one at a time, so to speak: it constructs an extended Hilbert
space $\mathcal{H}^+$ that depends on $\mathcal{M}(\Omega)$ but is
independent of $\beta$; then $\mathcal{H}$ is mapped isomorphically
onto a subspace that I denote $\tilde{\mathcal{H}}_\beta
\subset
\mathcal{H}^+$ that depends on $\beta$; implicit in the proof is an
isomorphism carrying each density operator $\rho \in
\mathcal{B}(\mathcal{H})$ to $\tilde{\rho}_\beta \in
\mathcal{B}(\tilde{\mathcal{H}}_\beta)$; this works in such a way that
$E^+$ is independent of $\beta$.  We want to swap these dependencies,
to make $\tilde{\mathcal{H}}$ independent of $\beta$ in exchange for
allowing a $\beta$-dependent $E^+_\beta$.  I claim this works as
follows. Pick any value of $\beta$, and call it 0; for this value of
$\beta$, $\mathcal{H}$ is mapped isomorphically onto
$\tilde{\mathcal{H}}_0$.  For any value of $\beta$ there exists a
unitary transform $U^+_\beta \in \mathcal{B}(\mathcal{H}^+)$, such
that \(\tilde{\rho}_\beta \oplus \tilde{\bm{0}}^\perp_\beta =
U_\beta^+(\rho \oplus \bm{0}^\perp)U^\dag_\beta.\) From this and
Proposition (\ref{eq:prop}) it follows that
\begin{eqnarray}
(\forall\ \beta, X)\quad \mbox{Tr}[\rho M_\beta(X)] & = &
\mbox{Tr}^+[E^+(X)(\tilde{\rho}_\beta + \tilde{\bm{0}}^\perp_\beta)]
\nonumber \\ &
= & \mbox{Tr}^+[U_\beta^+(\rho +
\bm{0}^\perp)U^{+\dag}_\beta E^+(X)] \nonumber \\
& = & \mbox{Tr}^+[(\rho +
\bm{0}^\perp)U^{+\dag}_\beta E^+(X)U_\beta^+] \nonumber \\ & = &
\mbox{Tr}^+[(\rho +
\bm{0}^\perp)E^+_\beta(X)],
\label{eq:help}
\end{eqnarray}
where we define \( E^+_\beta(X)\stackrel{\rm def}{=} U^{+\dag}_\beta
E^+(X)U_\beta^+ \).  Equation (\ref{eq:help}) and Lemma
(\ref{eq:lemProp}) imply that for an orthogonal projection $Q$ onto
$\tilde{\mathcal{H}}$, independent of $\beta$,
\( (\forall\ \beta,
X,\rho)\quad\mbox{Tr}[\rho M_\beta(X)]  =  \mbox{Tr}[\rho
(QE^+_\beta(X)Q)_{\tilde{\mathcal{H}}}],\) from which the conclusion of
the corollary follows. $\Box$

A simple example in finite dimensions occurs in \cite{MB97}, where a
plane on which a POVM is defined is embedded in a $\beta$-dependent
way in a three space equipped with a fixed projection-valued measure
defined by three mutually orthogonal basis vectors.  As in the
corollary, one can just as well hold the plane fixed and rotate the
basis vectors.

We apply this as follows.  Consider a family of
POVMs indexed by $\alpha$,
$M_{A,\alpha}:\mathcal{M}(\Omega_A)\rightarrow
\mathcal{B}(\mathcal{H}_A)$, along with another family of POVMs
indexed by $\beta$, $M_{B,\beta}:\mathcal{M}(\Omega_B)\rightarrow
\mathcal{B}(\mathcal{H}_B)$.  By the corollary, we translate Diagram 5
into a diagram for these families.  For all $\alpha$, $\beta$
we\vadjust{\kern6pt} have:
\begin{equation}
{\def\arraystretch{1.9}
\begin{tabular}{ccccccccccc}
&$E^+_{A,\alpha}$ &$\rightarrow$ & $E^+_{A,\alpha}\otimes \bm{1}_B$ &
$\rightarrow$ & $E^+_{A,\alpha}\otimes E^+_{B,\alpha}$ &$\leftarrow$ &
$\bm{1}_A \otimes E^+_{B,\beta}$ & $\leftarrow$ & $E^+_{B,\alpha}$ &
\\ $Q_A,|_{\mathcal{H}_A}$&$\downarrow$ && $\downarrow$ && $\downarrow$
&&
$\downarrow$ && $\downarrow$ & $Q_B,|_{\mathcal{H}_B}$\\ &
$M_{A,\alpha}$ &$\rightarrow$ & $M_{A,\alpha}\otimes \bm{1}_B$ &
$\rightarrow$ &
$M_{A,\alpha}\otimes M_{B,\beta}$ &$\leftarrow$ & $\bm{1}_A \otimes
M_{B,\beta}$ & $\leftarrow$ & $M_{B,\beta}$ &
\label{eq:diag6}
\end{tabular}\ .}
\end{equation}
\begin{center}
\vskip6pt
Diagram 6: Tensor product of POVMs of two families.
\end{center}
By the corollary, we have a diagram in which $Q_A$ and $Q_B$ are
independent of the indices $\alpha$, $\beta$.

\section{Applying quantum probabilities to descriptions of
devices}\label{sec:4}

Designs for systems of devices often start with models expressed in
simplified equations; quantum cryptography is a case in point.
Implementing a design inspired by equations entails arranging
devices---lasers, detectors, counters---so that measured device
behavior accords with properties expressed in the equations.
Experience teaches that the devices work as desired only when nudged
in ways unexpressed by the starting equations.  To support this
nudging, one ends up with layers of more detailed equations, needed
for instance in order to design feedback loops that compensate for
various drifts.  Whatever details we undertake to model, we face
choices.  For example, in an experiment with pulsed light, if we
assume there are no memory effects in the light detectors, we can
implement the state preparation for a trial by generating a single
light pulse, while to study memory effects in detectors, we must
implement the state preparation for a trial by generating a sequence of
pulses \cite{ams}.  When the equations used in modeling are equations in
the language of quantum mechanics, different choices are expressed by
different probability distributions stemming from different density
operators and different positive operator-valued measures (POVMs),
along with different choices of a measure space.\looseness=-1

\subsection{Choice of measure space}

To use a measure space as part of a quantum model of devices, one
needs, somehow, to link parameters---frequencies, positions, times---by
which one speaks in the laboratory of light pulses to the measure
space.  Several things complicate this linking.  First, no
single-frequency light state is an element of any separable Hilbert
space.  Second are laboratory facts (filters spill over, light
diffracts, signal times are fuzzy).  In modeling these facts, we need
detection operators that are correspondingly unsharp in relation to
the same parameters.  Third, the specification of a measure space in
terms of physical parameters depends on the layer of detail, and this
impedes comparisons of models across different layers of detail.
These complications are eased by adapting a trick from quantum
decision theory, in which POVMs are defined not with respect to an
arbitrary measure space but only for a discrete measure space, the
elements of which are thought of not in terms of physical parameters
but as possible actions to take in response to a measurement event.

So far the measure spaces appearing in this report have been
arbitrary, i.e., without any additional constraints.  They allow for
real metric spaces needed to deal with continuous spectra of hermitian
detection operators, and these measure spaces necessarily involve the
physical parameters of the spectrum.  We can preserve the capacity to
deal with these parameters when we must, while sidestepping their
complications when we can: the trick is to link arbitrary measure
spaces to discrete measure spaces adapted from quantum decision theory
\cite{helstrom,peres}.  In quantum decision theory, a POVM is defined
not with reference to a general measure space but as a countable set
of detection operators $M(j)$ that sum to $\bm{1}$.  One can imagine
the measurement result displayed by lighting up just one of a row of
lamps, indexed by $j$.  The context is one of action: ``if light $j$
goes on, do $X_j$.''  To link this to a general measure space
$\Omega$, we model an act of measuring not by a full POVM, but by a
countable set of detection operators $M(\Omega_j)$, where for $j = 1,
2,\ldots,$ the $\Omega_j$ are mutually disjoint subsets that cover
$\Omega$.  When we have in mind a mapping from natural numbers to
measurable sets of some measure space $\Omega$, we can abbreviate
$M(\Omega_j)$ by $M(j)$.  This abbreviation establishes a relation
among detection operators at different levels of detail with distinct
measure spaces $\Omega$ and $\Omega'$ by relating say $M(\Omega_j)$
and $M(\Omega'_j)$ to the same decision-oriented event $j$.  By this
linking of any arbitrary measure space to the natural numbers as a
single discrete measure space, we make universally applicable the
notion of a family of POVMs defined on that discrete measure space.

\section{Two applications of conditional probabilities}\label{sec:5}

Here are two applications of conditional probabilities as discussed
above, the first conceptual, the second concrete. 

\subsection{Sequences of probes in place of ``consecutive measurements''}

Tensor-product spaces are well suited to the modeling of the
interaction of a particle with a succession of probes, followed by
measurement of the probes, and in this connection a variety of
conditional density operators can be useful.  For example, express the
particle to be probed by a density operator $\rho_0 \in
\mathcal{B}(\mathcal{H}_0)$, and express probe $j$, $j = 1, 2,\dots,$
prior to its interaction with the particle by a density operator
$\rho_j \in \mathcal{B}(\mathcal{H}_j)$.  After a succession of
interactions (shown in Fig.~\ref{fig:1}), expressed by unitary operators
$U_{0j}
\in \mathcal{B}(\mathcal{H}_0\otimes\mathcal{H}_j)$, the probes are
measured, as expressed by POVMs $M_j:\mathcal{M}(\Omega_j) \rightarrow
\mathcal{B}(\mathcal{H}_j)$ on the respective Hilbert spaces
$\mathcal{H}_j$.  For two probes, this procedure yields for the joint
probability of components $X_j$ of the measurement
event:
\begin{figure}[t]  
\begin{center}
\includegraphics[width=3.5in]{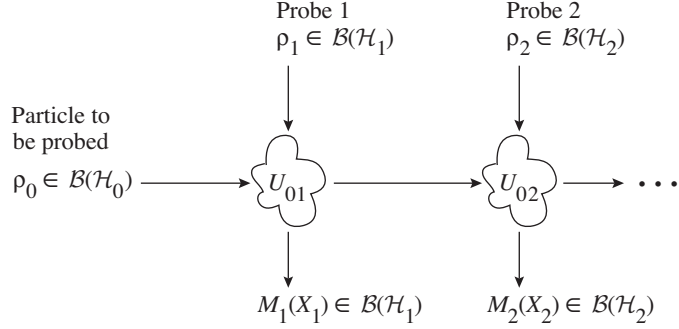}\vadjust{\kern-10pt} 
\end{center}
\caption{Particle $\rho_0$ undergoing successive 
interactions with probes $\rho_1$, $\rho_2$, \ldots}\label{fig:1}
\end{figure}
\begin{eqnarray}
\Pr(X_1,X_2) & = &
\mbox{Tr}_{012}\left(M_2(X_2)M_1(X_1)U_{02}\{[U_{01}(\rho_0\otimes\rho_1)
U_{01}^\dag]\otimes \rho_2\}U_{02}^\dag\right) \nonumber \\ & = &
\mbox{Tr}_{02}[M_2(X_2)U_{02}(\sigma_{0}|_{X_1}\otimes\rho_2)U_{02}^\dag],
\end{eqnarray}
where\vadjust{\kern-8pt}
\begin{equation}
\sigma_{0}|_{X_1} = \mbox{Tr}_1[M(X_1)U_{01}(\rho_0\otimes\rho_1)
U_{01}^\dag].\label{eq:s0x1}
\end{equation}
The corresponding conditional probability of $X_2$ given $X_1$ is
\begin{equation}
\Pr(X_2|X_1) =
\mbox{Tr}_{02}[M_2(X_2)U_{02}(\rho_{0}|_{X_1}\otimes\rho_2)U_{02}^\dag],
\end{equation}
where $\rho_{0}|_{X_1}$ is the (normalized) conditional density
operator engendered by the scaled conditional density operator
$\sigma_{0}|_{X_1}$:
\begin{equation}
\rho_{0}|_{X_1}
\stackrel{\rm def}{=} \frac{\sigma_{0}|_{X_1}}{\mbox{Tr}_{02}\
\sigma_{0}|_{X_1}}.
\end{equation}

For three probes, this procedure yields for the joint
probability of components $X_j$ of the measurement event:
\begin{eqnarray}
\Pr(X_1,X_2,X_3) & = &
 \mbox{Tr}_{023}\left[M_3(X_3)M_2(X_2)U_{03}\left(
[U_{02}(\sigma_0|_{X_1}\otimes\rho_2)U_{02}^\dag]\otimes
\rho_3\right)U_{03}^\dag\right] \nonumber \\ & = &
\mbox{Tr}_{03}[M_3(X_3)U_{03}(\sigma_0|_{X_1,X_2}\otimes\rho_3)
U_{03}^\dag],
\end{eqnarray}
where $\sigma_0|_{X_1}$ is given in Eq.\ (\ref{eq:s0x1}) and
\begin{eqnarray}
\sigma_0|_{X_1,X_2} & = &
\mbox{Tr}_2[M(X_2)U_{02}(\sigma_{0}|_{X_1}\otimes\rho_2) U_{02}^\dag] .
\end{eqnarray}
Corresponding to these equations, we find for the conditional
probabilities
\begin{eqnarray}
\Pr(X_2,X_3|X_1)& = & \mbox{Tr}_{023}\left[M_3(X_3)M_2(X_2)U_{03}\left(
[U_{02}(\rho_0|_{X_1}\otimes\rho_2)U_{02}^\dag]\otimes
\rho_3\right)U_{03}^\dag\right],\quad \\ \Pr(X_3|X_1,X_2) & = &
\mbox{Tr}_{03}[M_3(X_3)U_{03}(\rho_0|_{X_1,X_2}\otimes\rho_3)
U_{03}^\dag],
\end{eqnarray}
where each conditional density operator is defined as usual by
dividing a corresponding scaled conditional density operator by its
trace.

Recognizing freedom to invoke probes such as these in modeling
measurements has been shown to make the notion of repeated
measurements unnecessary in formulating the mathematics of quantum
mechanics in terms of projection-valued measures: a pair of successive
measurements can be subsumed into a single measurement involving a
succession of interactions of probes \cite{0404113}.  By virtue of
Neumark's theorem and the corollary above which makes it applicable to
a family of POVMs, this now generalizes to POVMs.  Once freedom to
invoke probes is accepted, there is no place nor any need in the logic
of quantum mechanics for a postulate pertaining to consecutive
measurements.  In particular, notwithstanding the efforts of Dirac
\cite{dirac} and von Neumann \cite{vN} and others
\cite{luders,davies} concerning consecutive measurements, there is no
place for a postulate of so-called ``state
reductions.''

\subsection{Example from quantum cryptography} 
\begin{figure}[t]  
\begin{center}
\includegraphics[width=4in]{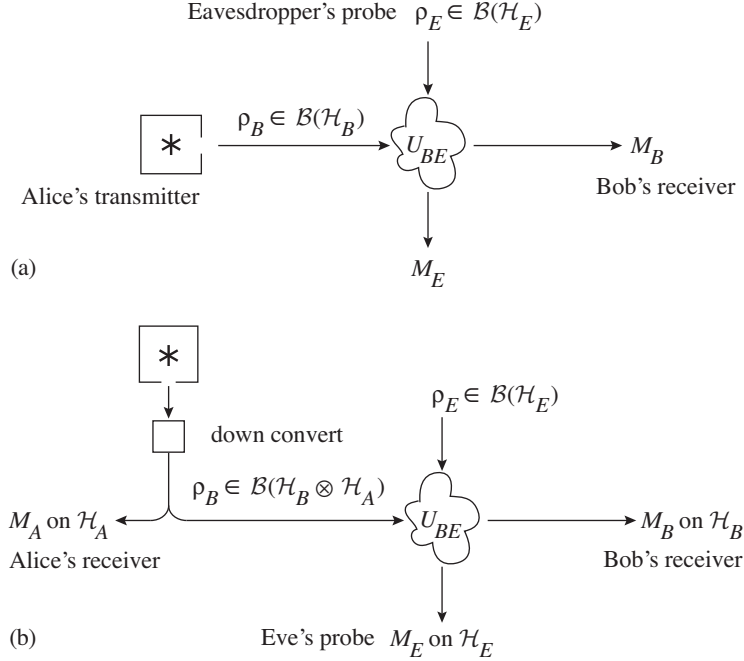}\vadjust{\kern-10pt}
\end{center}
\caption{(a) Transmitted-state QKD.\ \  (b) Entangled-state
QKD.}\label{fig:2}
\end{figure}

Models of quantum key
distribution (QKD) subject to individual eavesdropping attacks
\cite{qkd,slutsky} posit a sequence of trials, one trial for
each raw key bit.  The popular key protocol BB84
\cite{BB84} has two versions, ``transmitted-state'' (Fig.\
\ref{fig:2}a) and ``entangled-state'' (Fig.\ \ref{fig:2}b).  
Transmitted-state BB84 calls at each trial for Alice to prepare at
random one of four light states
$\rho_B(i)$, $i = 1,\ldots,4,$ with prior probabilities $\zeta_i$; these
states, subject to probing by Eve, are detected by Bob.  Entangled-state
BB84 calls for Alice to prepare a single polarization-entangled light
state
$\rho_{BA}$ that propagates to both her own detectors and to Bob's
detectors, again with the propagation to Bob subject to probing by Eve. 
Here I develop a relation that enables one to formulate individual
attacks against the seemingly more complicated entangled-state BB84 in
the same way as individual attacks against transmitted-state
BB84.\looseness=-1

The formulation by Slutsky {\em et al}.\ \cite{slutsky} for
transmitted-state BB84 holds for general light states defined in
\cite{jm05,Frwk}, not just the simplified states for which they
carried through their analysis.  As discussed in detail in
\cite{jm05,SPIE}, all the probabilities pertinent to key distribution
in the face of an individual eavesdropping attack against
transmitted-state BB84 stem from $\zeta_i$ together with the trace
rule applied to a quantum state on a tensor-product space
$\mathcal{H}_E \otimes \mathcal{H}_B$, where $\mathcal{H}_B$ is a
Hilbert space of light modes transmitted to Bob and $\mathcal{H}_E$ is
the Hilbert space of Eve's probe.  These probabilities have the
form
\begin{equation}
\Pr(X_E,Y_B,\rho_B(i)) =
\zeta_i\mbox{Tr}_{EB}[M_E(X_E)M_B(Y_B)U_{EB}(\rho_E\otimes
  \rho_B)U^\dag_{EB}],
\label{eq:trans}
\end{equation}
where $U_{EB}$ represents an arbitrary unitary interaction chosen by
Eve between her probe $\rho_E$ and Bob's light state $\rho_B$.\goodbreak

In models of entangled-state BB84, there are no prior probabilities
$\zeta_i$; instead, the Hilbert space involves three factors
$\mathcal{H}_E \otimes \mathcal{H}_B \otimes \mathcal{H}_A$, where
$\mathcal{H}_A$ is the Hilbert space for light detected by Alice.  At
each trial Eve prepares a probe $\rho_E$ as before, but now Alice
prepares an entangled state $\rho_{BA} \in \mathcal{B}(\mathcal{H}_B
\otimes \mathcal{H}_A)$.  With $U_{EB}$ as in transmitted-state BB84,
the probabilities pertinent to key distribution in the face of an
individual eavesdropping attack against entangled-state BB84 are just
\begin{equation}
\Pr(X_E,Y_B,Z_A) =
  \mbox{Tr}_{EBA}[M_E(X_E)M_B(Y_B)M_A(Z_A)U_{EB}(\rho_E\otimes
  \rho_{BA})U^\dag_{EB}],
\label{eq:tang}
\end{equation}
which looks significantly different from Eq.\ (\ref{eq:trans});
however, by the partial trace manipulations of Appendix A, in 
particular Eq.\ (\ref{eq:qkd}), this last equation
becomes
\begin{equation}
\Pr(X_E,Y_B,Z_A)  = 
   \mbox{Tr}_{EB}\{M_E(X_E)M_B(Y_B)U_{EB}[\rho_E\otimes
  \mbox{Tr}_A(M_A(Z_A)\rho_{BA})]U^\dag_{EB}\}. 
\end{equation}
This implies the following probability for Alice's component $Z_A$ of a
measurement event (regardless of Bob's and Eve's):
\begin{eqnarray}
\Pr(Z_A) & = & \mbox{Tr}_{EB}\{U_{EB}[\rho_E\otimes
  \mbox{Tr}_A(M_A(Z_A)\rho_{BA})]U^\dag_{EB}\} \nonumber \\ & = &
  \mbox{Tr}_{EB}[\rho_E\otimes \mbox{Tr}_A(M_A(Z_A)\rho_{BA})]
  \nonumber \\ & = & \mbox{Tr}_{B}[\mbox{Tr}_A(M_A(Z_A)\rho_{BA})],
\label{eq:zetTang}
\end{eqnarray}
which takes the part of $\zeta_i$ in transmitted-state BB84.
Taking the part of Eq.\ (\ref{eq:trans}) is the conditional
probability
\begin{eqnarray}
\Pr(X_E,Y_B|Z_A)& \stackrel{\rm def}{=}
&\frac{\Pr(X_E,Y_B,Z_A)}{\Pr(Z_A)} \nonumber \\ & = &
\frac{\mbox{Tr}_{EB}\{M_E(X_E)M_B(Y_B)U_{EB}[\rho_E\otimes
\mbox{Tr}_A(M_A(Z_A)\rho_{BA})]U^\dag_{EB}\}}
{\mbox{Tr}_{B}[\mbox{Tr}_A(M_A(Z_A)\rho_{BA})]} \nonumber \\
& = &\mbox{Tr}_{EB}[M_E(X_E)M_B(Y_B)U_{EB}(\rho_E\otimes
\rho_B|_{Z_A})U^\dag_{EB}],
\label{eq:tang2} 
\end{eqnarray}
with the conditional density operator
\begin{equation} 
\rho_{B}|_{Z_A} \stackrel{\rm def}{=}\frac{
\mbox{Tr}_A[M_A(Z_A)\rho_{BA}]}
{\mbox{Tr}_{B}[\mbox{Tr}_A(M_A(Z_A)\rho_{BA})]}.
\label{eq:qkd2}
\end{equation}  Entangled-state BB84 requires that the
state-event components $Z_A$ for Alice include four that
correspond to the four choices Alice has in transmitted-state
BB84.  I call these $Z_{A,i}$. 
By virtue of Eqs.\ (\ref{eq:zetTang}) and (\ref{eq:tang2}) we
arrive at the following

\vspace{5pt}

\noindent\textbf{Proposition}: Any model of entangled-state BB84 of
the form of Eq.\ (\ref{eq:tang}) asserts the same joint probabilities
relevant to quantum key distribution subject to individual
eavesdropping attacks as does the model of transmitted-state BB84 with
$\zeta_i = \Pr(Z_{A,i})$ defined by Eq.\ (\ref{eq:zetTang}) and with
$\rho_B(i) = \rho_{B}|_{Z_{A,i}}$ defined by Eq.\ (\ref{eq:qkd2}).

\acknowledgments
For very helpful discussions over the past four years, I thank Tai
Tsun Wu.  This work was supported in part by the Air Force Research
Laboratory and DARPA under Contract No. F30602-01-C-0170 with BBN
Technologies.

\appendix
\section{Tensor products and partial traces}\label{app:A}

By a bounded positive operator $A$ on a Hilbert space $\mathcal{H}$, I
mean any bounded self-adjoint operator such that $(\forall\ |x\rangle
\in \mathcal{H})\ \ \langle x|A|x\rangle \ge 0$.  The trace of a
positive operator $A$ on a separable Hilbert space is defined
\cite{sewell} as
\begin{equation}
\mbox{Tr}(A) = \sum_n \langle \psi_n|A|\psi_n\rangle,
\end{equation} where $\{\psi_n\}$
is an orthonormal basis.  In fact, the value of $\mbox{Tr}(A)$ here,
which might be infinite, is independent of the choice of basis
\cite{vN}.  As stated in \cite{sewell} and proved in \cite{vN}, the
trace is invariant under cyclic permutations of the factors of a
product:
\begin{equation}   \mbox{Tr}(ABC)=\mbox{Tr}(CAB). \end{equation}

For a finite-dimensional vector space, the following discussion is
elementary; a start at the more complicated derivations needed for
infinite-dimensional, separable Hilbert spaces can be found in Chap.\
II, Sec.~11 of \cite{vN}.

The trace of a product of matrices acting on
a finite-dimensional vector space $\mathcal{H}$ is defined by
\begin{equation}
\mbox{Tr}(MN) = \sum_{j,k} M_{j,k}N_{k,j} = \sum_{j,k} N_{j,k}M_{k,j}
= \mbox{Tr}(NM).
\end{equation}
Suppose that $\mathcal{H} = \mathcal{H}_A\otimes\mathcal{H}_B$.
For the tensor product of two vectors $a \in \mathcal{H}_A$ and $b \in
\mathcal{H}_B$, with components $a_{J}$ and $b_{j}$, respectively, we
write the tensor product as a vector $w \in
\mathcal{H}_A\otimes\mathcal{H}_B$ with components $w_{Jj} =
a_{J}b_{j}$.  Now let $M^{(AB)}$ be a matrix operating on vectors of
$\mathcal{H}_A\otimes\mathcal{H}_B$.  A row of such a matrix is
specified by a double index such as $Jj$, and a column by a double
index such as $Kk$, so that in the special case $M^{(AB)}=
R^{(A)}\otimes S^{(B)}$, $M^{(AB)}_{JjKk} = R^{(A)}_{JK}S^{(B)}_{jk}$
(where we have assumed $R^{(A)}$ is a matrix acting on vectors of
$\mathcal{H}_A$ and $S^{(B)}$ is a matrix acting on vectors of
$\mathcal{H}_B$). The partial trace over the $\mathcal{H}_B$ factor of
$M^{(AB)}$ is defined to be the matrix acting on $\mathcal{H}_A$ that
has as its $(J,K)$-th component
\begin{equation} 
[\mbox{Tr}_B(M^{(AB)})]_{JK}= \sum_{j}M_{JjKj}.  \end{equation} For
the full trace over $\mathcal{H}_A\otimes\mathcal{H}_B$ I write
$\mbox{Tr}_{AB}$.

Any matrix $M^{(AB)}$ can be written as a sum of tensor products, with
the result that the properties of partial traces follow from block
form of these tensor-product terms.  Recall that for
finite-dimensional spaces the tensor product of any two matrices $A$
and $B$ is a matrix of $B$-sized blocks:
\begin{equation}
A\otimes B = \left[\begin{array}{lll}
A_{11}B & A_{12}B & \ldots \\
A_{21}B & A_{22}B & \ldots \\
\ldots & \ldots & \ldots
\end{array}\right] ,
\end{equation}
from which it is obvious that for square matrices $A$ and $B$,
\begin{equation}
\mbox{Tr}_B(A \otimes B) = \left[\begin{array}{lll}
A_{11}\mbox{Tr}(B) & A_{12}\mbox{Tr}(B) & \ldots \\
A_{21}\mbox{Tr}(B) & A_{22}\mbox{Tr}(B) & \ldots \\
\ldots & \ldots & \ldots
\end{array}\right] = [\mbox{Tr}(B)]A.
\end{equation}
Similarly, we have $\mbox{Tr}_A(A\otimes B) = [\mbox{Tr}(A)]B$,
from which it follows immediately that $\mbox{Tr}_{AB}(A\otimes B) =
[\mbox{Tr}_A(A)][\mbox{Tr}_B(B)]$.
Except when computing with matrix components, I drop
the superscripts indicating ``which space'' to subscripts, writing
$M_{AB}$ in place of $M^{(AB)}$, etc. 

Although the full trace of a product is invariant under a change in
the order of the factors, the block form makes apparent that changing the
order of factors affects the partial trace of their product; in the
general case,
\begin{eqnarray}
\mbox{Tr}_B[(R_A \otimes S_B)(R'_A\otimes S'_B)] & = &
\mbox{Tr}_B[(R_AR'_A)\otimes (S_BS'_B)] = [\mbox{Tr}(S_BS'_B)]R_AR'_A
\nonumber \\ & \bm{\not \equiv} & \mbox{Tr}_B[(R'_A \otimes
S'_B)(R_A\otimes S_B)].
\label{eq:noncom}
\end{eqnarray} 
It is of course true that for any
two matrices $M$ and $N$ acting on
$\mathcal{H}_A\otimes\mathcal{H}_B$, we have
\begin{eqnarray}
(\forall\ M, N)\quad \mbox{Tr}_A[\mbox{Tr}_B(MN)] & = &
\mbox{Tr}_A[\mbox{Tr}_B(NM)]\nonumber \\ & =
&\mbox{Tr}_B[\mbox{Tr}_A(MN)] = \mbox{Tr}_B[\mbox{Tr}_A(NM)] =
\mbox{Tr}_{AB}(MN).\qquad
\end{eqnarray}
Easily proved are the following facts concerning square matrices,
stated with subscripts to indicate the relevant vector spaces,
$\mathcal{H}_A$,
$\mathcal{H}_B$, and $\mathcal{H}_A\otimes\mathcal{H}_B$.  First,
there holds a commutativity relation for partial traces of factors
of a certain form:

\vspace{8pt}
\noindent\textbf{Lemma}:\vadjust{\kern-6pt} 
\begin{equation}
(\forall\ S_B,
M_{AB})\quad\mbox{Tr}_B[M_{AB}(\mathbf{1}_{A}\otimes S_B)] = 
\mbox{Tr}_B[(\mathbf{1}_{A}\otimes S_B)M_{AB}].
\label{eq:a9}
\end{equation}

\vspace{5pt}
\noindent\textit{Proof}: Expand any $M_{AB}$ as a sum of tensor
products $R_A \otimes S_B$, and observe that for this special case the
$\not \equiv$ of Eq.\ (\ref{eq:noncom}) becomes equality.
$\Box$

\vspace{8pt}

The same technique shows 

\noindent\textbf{Lemma}:\vadjust{\kern-6pt} 
\begin{equation}
(\forall\ \mbox{ hermitian }S_B, \mbox{ hermitian }M_{AB})\quad 
\mbox{Tr}_B[M_{AB}(\bm{1}_{A}\otimes S_B)] \mbox{ is hermitian.}
\label{eq:a10}
\end{equation}

\vspace{8pt}

Again the same technique shows

\noindent\textbf{Lemma}:\vadjust{\kern-6pt} 
\begin{eqnarray}
\mbox{Tr}_B[(R_{A}\otimes \bm{1}_B)M_{AB}] \mcl&=&\mcl
R_{A}\mbox{Tr}_B(M_{AB}),
\label{eq:a11}\\
\mbox{Tr}_B[M_{AB}(R_A\otimes \bm{1}_B)]\mcl&=&\mcl
[\mbox{Tr}_B(M_{AB})]R_A,\label{eq:a12}
\end{eqnarray}
from which follows 

\vspace{8pt}

\noindent\textbf{Lemma}:\vadjust{\kern-6pt} 
\begin{equation}
(\forall\ R_{A}, M_{AB})\quad\mbox{Tr}_{AB}[(R_{A}\otimes
\bm{1}_B)M_{AB}] = \mbox{Tr}_A[R_{A} \mbox{Tr}_B(M_{AB})].
\end{equation}
Note the stretch of notation so that $\mbox{Tr}_A$ is
used here for the trace on $\mathcal{H}_A$, while it is also used,
as above, for the partial trace over the $A$ factor of 
$\mathcal{H}_A\otimes\mathcal{H}_B$.
All these lemmas holds when $A$ and $B$ are interchanged; for instance
the last lemma becomes:
\begin{equation}
(\forall\ S_B, M_{AB})\quad\mbox{Tr}_{AB}[(\bm{1}_{A}\otimes
S_B)M_{AB}] = \mbox{Tr}_B[S_B \mbox{Tr}_A(M_{AB})].
\end{equation}

Here is an example of the application to quantum mechanics.
Sometimes one wants to compute a conditional density operator
of the form 
\begin{eqnarray}
\rho_A &=&
\mbox{Tr}_A\left(M_{A}(X)|\psi_{AB}\rangle\langle\psi_{AB}|\right)
\equiv \mbox{Tr}_A\left((M_{A}(X)\otimes \bm{1}_B)
|\psi_{AB}\rangle\langle\psi_{AB}|\right),
\label{eq:examp}
\end{eqnarray}
where $M_{A}(X)$ is positive hermitian and thus can be
written as a product of its positive square roots
$[M_{A}(X)]^{1/2}$. In some cases the calculation is made easier by
putting Eq.\ (\ref{eq:examp}) in a symmetric form.  To this end, 
interchange $A$ and $B$ in Lemma (\ref{eq:a9}) and change names to
obtain\looseness=-1

\vspace{5pt}
\noindent\textbf{Lemma}: 
\begin{equation} 
(\forall\ R_{A}, Q_{AB})\quad \mbox{Tr}_A[(R_{A}\otimes
\mathbf{1}_B)Q_{AB}] = \mbox{Tr}_A[Q_{AB}(R_{A}\otimes
\mathbf{1}_B)].
\label{eq:swap2}
\end{equation}
Apply this with $R_{A}$ taken as $[M_{A}(X)]^{1/2}$ and with
$Q_{AB}$ taken as
$[M_{A}(X)]^{1/2}|\psi_{AB}\rangle\langle\psi_{AB}|$ to obtain
the symmetrized form
\begin{eqnarray}
\rho_A &=&
\mbox{Tr}_A\left([M_A(X)]^{1/2}|\psi_{AB}\rangle
\langle\psi_{AB}|[M_A(X)]^{1/2}\right).
\end{eqnarray}

To deal with partial traces of more complex expressions, a couple
of tricks help.  The first is a shorthand notation.
So far I have written out lots of tensor products of
operators in which the operator on one of the factor spaces is the
identity operator for that space, as in
$\mbox{Tr}_B[M_{AB}(\bm{1}_A\otimes S_B)]$, which contains an operator
product of the operator $M_{AB}$ on $\mathcal{H}_A\otimes
\mathcal{H}_B$ times the indicated tensor product $\bm{1}_A\otimes S_B$.
Common in the physics literature and often convenient is a shorthand
convention of writing just $\mbox{Tr}_B(M_{AB}S_B)$.  This shorthand
can be undone by tensoring each operator in such an expression into 
the identity operators required for the whole expression to make sense.

The second trick is merely to recognize that if $\mathcal{H}_A$ and
$\mathcal{H}_B$ are distinct factor spaces of a tensor-product space
$\mathcal{H}_{AB}$, then 
\begin{equation}
[M_A,N_B] = [(M_A \otimes \bm{1}_B)(\bm{1}_A\otimes N_B)] = 0;
\end{equation}
that is, operators on distinct factor spaces commute with one another.
For example, Lemmas (\ref{eq:a9}) and (\ref{eq:a10}) with
a swapping of $A$ and $B$ become in shorthand notation
\begin{eqnarray}
\mbox{Tr}_A(M_{AB}R_A) & = & \mbox{Tr}_A(R_AM_{AB}) \\
\mbox{Tr}_A(S_BM_{AB})& = & S_B\mbox{Tr}_A(M_{AB}).
\end{eqnarray}
{}From these equations follows a more complicated relation for
operators on a triple tensor product $\mathcal{H}_E\otimes
\mathcal{H}_B\otimes\mathcal{H}_A$, of
use in quantum cryptography:
\begin{eqnarray}
\mbox{Tr}_A[M_EM_BM_AU_{EB}\rho_E\rho_{AB}U^\dag_{EB}]
& = & \mbox{Tr}_A[M_EM_BU_{EB}\rho_EM_A\rho_{AB}U^\dag_{EB}]  \nonumber \\
& =  &  M_EM_BU_{EB}\rho_E[\mbox{Tr}_A(M_A\rho_{AB})]U^\dag_{EB}.
\label{eq:qkd}
\end{eqnarray}

\end{document}